# The challenges of changing teaching assistants' grading practices: Requiring students to show evidence of understanding

Emily Marshman[1], Ryan Sayer[2], Charles Henderson[3],
Edit Yerushalmi[4] and Chandralekha Singh[1]

[1] *Department of Physics and Astronomy, University of Pittsburgh, 3941 O'Hara St., Pittsburgh, PA 15260 USA*
[2] *Department of Physics, Bemidji State University, Bemidji, MN 56601 USA*
[3]*Department of Physics, Western Michigan University, 1903 W. Michigan Ave., Kalamazoo, MI 49008 USA*
[4]*Department of Science Teaching, Weizmann Institute of Science, 234 Herzl St., Rehovot, Israel 7610001*

**Abstract:** Teaching assistants (TAs) are often responsible for grading in introductory physics courses at large research universities. Their grading practices can shape students' approaches to problem solving and learning. Physics education research recommends grading practices that encourage students to provide evidence of understanding via explication of the problem-solving process. However, TAs may not necessarily grade in a manner that encourages students to provide evidence of understanding in their solutions. Within the context of a semester-long TA professional development course, we investigated whether encouraging TAs to use a grading rubric that appropriately weights the problem-solving process and having them reflect upon the benefits of using such a rubric prompts TAs to require evidence of understanding in student solutions. We examined how the TAs graded realistic student solutions to introductory physics problems before they were provided a rubric, whether TAs used the rubric as intended, whether they were consistent in grading similar solutions, and how TAs' grading criteria changed after discussing the benefits of a well-designed rubric. We find that many TAs typically applied the rubric consistently when grading similar student solutions but did not require students to provide evidence of understanding. TAs' written responses, class discussions, and individual interviews suggest that the instructional activities involving the grading rubrics in this study were not sufficient to change their grading practices. Interviews and class discussions suggest that helping TAs value a rubric that appropriately weights the problem-solving process may be challenging partly due to the TAs' past educational experiences and the departmental context.

## I. INTRODUCTION

At large research institutions in the U.S., graduate students in physics play an important role in the education of undergraduate students in physics courses. In particular, it is quite common for physics graduate Teaching Assistants (TAs) to teach introductory physics recitations or lab sections. TAs are partly responsible for helping undergraduate students achieve the goals of introductory physics course instructors, which typically include learning disciplinary concepts and principles [1], developing effective problem-solving approaches, and making effective use of problem solving as an opportunity for learning [2-11].

One way to promote instructors' goals is via grading [12-16], for which TAs are often responsible. Physics education research recommends grading practices that encourage students to provide evidence of understanding via explication of the problem-solving process, which can help them learn physics concepts and develop effective problem-solving approaches [2-11]. However, TAs' grading practices are shaped partly by their past experiences as students and the departmental context, including expectations of their supervising faculty member [17-22]. Prior research suggests that many TAs may not have thought about the benefits of encouraging their students to show evidence of understanding and they may not require students to explicate the problem-solving process when grading their solutions [21,22].

Professional development courses can provide valuable opportunities for helping TAs contemplate how requiring their students to explicate the problem-solving process can promote the common goals and learning outcomes for introductory physics courses. The case study presented here investigated how a sequence of grading activities involving a grading rubric implemented in a mandatory, one-semester professional development course impacted TAs' grading practices. The cognitive apprenticeship framework [23] inspired the grading activities implemented in the professional development course to improve TAs' grading beliefs and practices. The cognitive apprenticeship framework involves "modeling" (i.e., the instructor demonstrates and exemplifies the skills that students should learn), "coaching and scaffolding" (i.e., providing students suitable practice, guidance, and feedback so that they learn the skills necessary for good performance), and "weaning" (i.e., gradually reducing the scaffolding support to help students develop self-reliance). A grading rubric was developed that weights the problem solving process appropriately to help students learn physics and develop expert-like problem solving practices. The rubric was used to model the criteria of good performance, and TAs were coached and provided scaffolding support on how to use the rubric and why consistently using the rubric to grade would help students develop expert-like problem solving strategies and learn physics. The TAs were asked to grade several realistic solutions (with different levels of explication of the problem-solving process) for two isomorphic introductory physics problems with and without the rubric. TAs also participated in small group and class discussions about the rubric to help them reflect on the advantages of using a rubric that requires students to explicate the problem solving process and appropriately weights effective problem-solving practices. Finally, TAs were "weaned" in that they were asked to grade solutions to physics problems again at the end of the professional development course without being given a rubric. We hypothesized that grading activities inspired by cognitive apprenticeship may prompt TAs to reflect upon how grading on specific criteria (that they may not have considered on their own) can encourage students to provide evidence of understanding, which can help students learn physics concepts and develop effective problem-solving approaches. The study was designed to investigate the following research questions:

1. *How do TAs apply the different components of the rubric that weights the problem-solving process to grade student solutions of introductory physics problems?*

2. *Do TAs use the rubric consistently when grading solutions to problems involving the same physics principles but having different surface features?*

3. *Do TAs apply the grading rubric differently than an "expert rater", e.g., physics education researchers who study problem solving?*

The study was also designed to examine whether TAs' grading practices changed after the professional development course and investigate TAs' perceptions about the rubric:

4. *Do TAs' grading practices change after a 15-week professional development course that included activities and extended discussions involving the benefits of using a rubric and carrying out their assigned teaching responsibilities simultaneously?*

5. *According to the TAs, what are the pros and cons of using a rubric to grade student solutions in introductory physics?*

The findings can inform the leaders of professional development courses for TAs and instructors and physics education researchers in contemplating strategies for improving beliefs about grading and grading practices to foster learning.

## II. BACKGROUND

### A. Effective problem-solving approaches

Many prior studies [2-11] have documented differences between novices and experts when approaching problems. Both use heuristics to guide their search process in identifying the gap between the problem goal and the state of the solution and taking action to bridge this gap. However, novices differ from experts in the types of heuristics they use to solve problems. Novices approach problems in a haphazard manner, typically searching for appropriate equations first and plugging in numbers until they get a numerical answer [2-8]. Furthermore, novices often draw on their naive knowledge base rather than formal physics knowledge [2-8]. Novices also engage in pattern matching, i.e., attempting to solve a problem using another previously solved problem with similar surface features, even if the underlying concepts and principles are different [2-8]. On the other hand, experts devote time and effort to qualitatively describe the problem situation, identify principles and concepts that may be useful in the analysis of the problem, and retrieve effective representations based on their better organized domain knowledge [2-8]. In addition, experts devote time to plan a strategy for constructing a solution by devising a useful set of intermediate goals and means to achieve them [2-8]. Experts also spend more time than novices in using diverse representations to analyze and explore problems (especially when they are not sure how to proceed) [2-8]. Experts also engage more than novices in self-monitoring by evaluating previous steps and revising their choices as needed [9-11]. They utilize problem solving as a learning opportunity more effectively by engaging in self-repair - identifying and attempting to resolve conflicts between their own mental model and the scientific model conveyed by peers' solutions or worked-out examples [9-11].

Effective grading practices can impact the development of expertise. For example, grading on the problem-solving process can impact whether students use effective approaches to problem solving (e.g., starting problem solving with a conceptual analysis of the problem, doing planning and decision making before implementing the plan, and then doing a reasonability check for the solution obtained and reflecting upon the problem-solving process to learn from it) instead of a plug and chug approach (e.g., starting by looking for a formula that matches the quantities in the problem statement). In a prior research study [4], students in two different groups were matched in terms of their prior performance in traditionally taught physics classes (for example, students who have a C grade at the time of the interview are matched with other students with a similar grade). Students in one of the two groups were required to use effective approaches to problem solving (experimental group) and those in the other group were allowed to use whatever approach they want to use to solve the problem (control group). The performance of the students in the experimental group was significantly better than those in the control group as the complexity of the problem increased [4]. Since students often value what they are graded on, grading their solutions on the explication of the problem-solving process can encourage students to use a systematic approach to problem solving.

### B. Grading rubrics

Findings of physics education research (PER) recommend grading practices that encourage students to show evidence of understanding via explication of the problem-solving process. In the spirit of formative assessment, effective grading practices can provide feedback that can foster learning, communicate to learners what practices are useful in learning the discipline and for developing problem solving skills [24-26]. Effective grading practices can also communicate to students what to focus on in the future learning activities [27-32]. Such practices can encourage students to explain the reasoning underlying their solutions and provide them with an artifact to reflect on and learn from after problem solving (i.e., their own clearly articulated solution in which the problem-solving process is explicated) [33].

Grading rubrics are scoring tools which outline the performance expectations for an assignment. Good rubrics often divide a problem into various parts and provide descriptions of how scores should be allocated for varying levels of mastery. Effective grading rubrics offer many advantages to both students and instructors. A grading rubric that rewards explication of the problem-solving process (instead of focusing mainly on the correctness of the final answer) can give students an incentive to use problem solving strategies which are useful for the development of important skills and learning physics. As noted earlier, when students are required to use effective problem-solving approaches, they are significantly more successful in solving complex physics problems compared to matched students with similar course grades at the time [4]. A good grading rubric can provide a consistent grading standard for students with a focus on approaches that enhance students' knowledge and skills. Research in various domains has shown that good rubrics can serve as formative assessment tools for students, helping them recognize strengths and weaknesses of their work and monitor their progress toward mastery [34-48]. By knowing ahead of time what is expected of their work (if they are given a rubric and informed that they will always be graded on it), students may be encouraged to practice effective problem-solving strategies (e.g., initial analysis and planning, explication, reasonability check for the solution and other types of reflection on the problem-solving process to learn from solving the problem) that may help them develop problem-solving skills and learn [34-48]. Students may also develop a better understanding of their difficulties if they are graded via a rubric and they may focus on developing a better knowledge structure and problem solving skills. In addition, students and instructors may have a more consistent judgment of the students' work because the use of rubrics decreases variation in scores between different graders [34-48].

An effective rubric can provide feedback to the students and the instructor on students' problem-solving approaches. Docktor et al. [39] designed a rubric to assess expertise in problem solving. Their rubrics assess students' proficiency related to both content knowledge and skills and include the processes of organizing problem information into a useful description, selecting appropriate physics principles, applying physics concepts and principles to the specific situations in the problem, using mathematical procedures appropriately, and communicating an organized reasoning pattern [39]. Such a rubric focuses on evidence of understanding, initial problem analysis, and evaluation as opposed to only correctness of the final answer and algebra. Other researchers have developed rubrics that involve similar components, e.g., describe the physics, plan the solution, execute the plan, and evaluate the solutions [41-43,47].

### C. Physics graduate TAs and their typical role and training

In introductory physics courses (both recitations and labs) at large research universities, graduate TAs are often responsible for grading homework and quizzes and often at least part of the exams (part of the exam may be graded by the course instructor). At the Graduate Education in Physics Conference jointly sponsored by the American Physical Society and the American Association of Physics Teachers, discussions with faculty members about teaching assistantships suggest that the majority of physics departments at research institutions in the U.S. employ physics graduate students as TAs for introductory physics course recitations and for introductory laboratory classes [49]. The TAs are expected to do the bulk of grading in these courses. A majority of physics departments provide a very short training to the TAs (half day or less) to help them learn how to carry out their teaching responsibilities [49]. However, a handful of departments have provided semester long TA professional development programs similar to the one discussed in this study. Moreover, most conference participants noted that the TAs usually carry out the tasks in their recitations, labs, and grading without significant supervision or guidance from their supervising instructor except for general guidelines about how to carry out the recitation or how to grade (e.g., whether they should solve homework problems on the board in the recitation, give a quiz at the beginning or at the end of the recitation, how easy or strict they should be in grading homework and quizzes, etc.) [49].

### D. Shifting physics graduate TAs' instructional beliefs and practices

Prior studies have identified common beliefs and practices among physics TAs that have implications for improving learning [18-22,50-55]. For example, research suggests that TAs sometimes struggle to understand the value of thinking about the difficulty of a problem from an introductory student's perspective [53,54] and believe that if they know the content and can explain it to their students in a clear manner, it will be sufficient to help their students learn. Also, while physics TAs are able to recognize useful solution features and articulate why they are important when looking at sample introductory physics student solutions provided to them, they do not necessarily include those features in their own solutions written for introductory physics courses [55].

We previously investigated TAs' beliefs about grading and implemented activities to improve their beliefs about grading in the context of a one-semester professional development course [21,22]. We found that many TAs understand that grading can help students learn physics, develop problem solving skills, and help instructors identify common student difficulties [21,22]. However, these same TAs gave a higher score to a solution that provided minimal reasoning while possibly obscuring physics mistakes as compared to a solution that provided reasoning but revealed canceling mistakes [21,22]. The solution features that the TAs graded on were generally correct application of physics knowledge and correct final answer—few TAs graded on problem description, evaluation, and explication of the problem solving approach. In the prior investigation, we attempted to help TAs resolve this conflict by providing opportunities to discuss their grading practices in small groups and resolve any disagreements in grading. The TAs shared their grading approaches with the entire class, during which the instructor gave feedback. At the end of the TA professional development course, the TAs were again asked to individually write an essay about the purpose of grading and grade four realistic student solutions. This end of semester activity allowed us to observe whether TAs' grading practices and beliefs had changed over the course of the TA professional development course and one semester of teaching experience. We found that these activities were not successful in making significant changes to TAs' grading practices, and TAs generally continued to grade on the correct application of physics and the correct final answer and did not take into account students' explication of the problem-solving approach [21,22]. We hypothesized that the support provided to the TAs in the previous study [21,22] was not extensive enough to improve their grading practices and help them require students to explicate the problem-solving process.

## III. THEORETICAL FRAMEWORK

This study presented here attempts to build on the findings of a previous study [21,22] and uses a field-tested cognitive apprenticeship framework [23] to investigate how a sequence of grading activities implemented in a one-semester professional development course impacted TAs' grading beliefs and practices. The cornerstones of the cognitive apprenticeship framework are emphasis on modeling the criteria of good performance for the learners, providing them appropriate coaching and scaffolding support, and gradually reducing the support given to allow students to develop self-reliance. The cognitive apprenticeship framework guided the design of the grading activities implemented in the professional development course to improve TAs' grading beliefs and practices. In particular, the "modeling" component of the cognitive apprenticeship framework refers to the instructor demonstrating and exemplifying the knowledge and skills that TAs should learn in order to develop productive beliefs about grading and grading practices. "Coaching and scaffolding" refer to providing the TAs with practice, guidance, and feedback so that they develop productive beliefs about grading and grading practices. Finally, "weaning" refers to allowing TAs to grade solutions to problems without the explicit guide of a rubric and support from the instructor and/or their peers.

In the professional development course, the grading rubric itself was a model of the criteria for how TAs should grade solutions to physics problems. Then, TAs were coached and provided scaffolding support on how to use the rubric and given opportunities to reflect upon why consistently using the rubric to grade would help their students develop expert-like problem solving strategies and learn physics. They were asked to grade several introductory physics solutions for two isomorphic problems with and without the rubric, with different levels of explication of the problem-solving process. Further coaching and scaffolding support was provided to the TAs to improve their grading beliefs and practices. They were also given opportunities to participate in small group and class discussions and reflect upon the value of the rubric that requires to show evidence of understanding. At the end of the professional development program, TAs were "weaned" in that they were asked to grade solutions to physics problems without being given a rubric explicitly (i.e., we wanted to determine how much they internalized from the rubric activities and whether they would generate a grading rubric on their own). We hypothesized that the cognitive apprenticeship-inspired grading activities in the professional development course may prompt the TAs to reflect upon how effective grading practices can encourage students to provide evidence of understanding and help them learn physics concepts and develop problem solving strategies. In addition, TAs may also reflect on how grading with a rubric may lead to greater objectivity, consistency, and repeatability when assigning scores to student work. The coaching and scaffolding support provided to the TAs in the professional development course began with an attempt to help them reflect upon and account for their "resources" for teaching and learning, which includes their prior knowledge [56]. The social context in which the learning takes place is also a key component within the cognitive apprenticeship framework. In this study, the social context is supported in that the TAs are all participating in a professional development course and are all part of the same subculture. They also have access to a variety of expert instructors and have opportunities to observe their peers, who have varying degrees of skills.

The first-year physics graduate TAs in the TA professional development course were concurrently teaching recitations or laboratory sections. They were asked to individually 1) write an essay about the purpose of grading; and 2) grade four realistic student solutions based upon actual students' work to elicit their initial ideas about grading before the rubric was introduced. The instructor also attempted to use TAs' productive beliefs about grading as a resource to facilitate positive discussions and coached the TAs to think about the benefits of grading using a rubric that focuses on the problem-solving process. Since the graduate TAs who participated in this study were also simultaneously teaching introductory physics recitations or introductory labs, it was hypothesized that those activities may provide additional scaffolding and synergistic benefits for what the TAs reflected upon and learned in the TA professional development course.

## IV. METHODOLOGY AND DATA COLLECTION

### A. Participants

In this investigation, we collected grading data from a mandatory, semester-long TA professional development course at a large research university in the U.S. led by one of the authors. A total of 15 first-year TAs were enrolled in the course. The course met for 2 hours each week for the entire semester and was designed to prepare them for their teaching responsibilities. The TAs had also attended a day long new teaching assistant workshop facilitated by the university, but this workshop was general and did not focus on discipline-specific issues in teaching and learning physics. The TAs in general were expected to do one hour of homework each week pertaining to the professional development course. The majority of the TAs were concurrently teaching recitations for introductory physics courses for the first time. A few TAs were also assigned to facilitate a laboratory section or grade students' work in various physics courses for the first time. A majority of the TAs also worked as tutors in a resource room where introductory students are assisted with any help they need with physics including their physics homework and laboratory reports. The participants consisted of a mix of domestic and international students originating from nations such as China, India, Turkey, etc. There were 4 female TAs and 11 male TAs.

### B. Data collection tools and artifacts

The data on TAs' goals for grading and grading practices were collected using a group administered interactive questionnaire (GAIQ), previously developed and validated by three of the authors (E.Y., C.H. and C.S.) in collaboration with two graduate student researchers in physics education for use with TAs [19,57]. This tool consists of a series of activities involving worksheets and artifacts that are designed to clarify a TA's ideas about helping students learn physics content and problem solving skills. One component of the GAIQ focuses on TAs' views about grading (which is the component that we discuss here). Each component of the GAIQ includes several stages, including: 1. A pre-class activity in which TAs complete a worksheet eliciting their initial ideas; 2. In class, TAs work in groups of three to answer the same questions as in the pre-class worksheet and then a whole class discussion takes place in which groups share their work; 3. TAs individually complete another worksheet in which they can reflect on their previous ideas and modify them.

The specific artifacts involving grading activities included two isomorphic introductory physics problems (Problems 1 and 2) and four student solutions. Problem 1 (see Figure 1) and Problem 2 (see Figure 3) were designed, validated and approved by four physics instructors who taught introductory physics courses at the University of Minnesota and were used on final exams. Each problem involves synthesis of the same important physics concepts and principles, is context-rich, and is difficult enough to require an average student to use an exploratory decision making process as opposed to an algorithmic procedure [2-11]. There were two student solutions to each problem. The student solutions were based upon actual students' common answers in the final exam and reflect differences between expert and novice problem solving from the research literature such as including a diagram describing the problem, explication of sub-problems, justification of steps, evaluation of the final answer, explication of the principles used, evidence of reflective practices, etc. [2-11]. Figure 2 shows Student Solution D (SSD) and Student Solution E (SSE), which were solutions to Problem 1. Figure 4 shows Student Solution F (SSF) and Student Solution G (SSG), which were solutions to Problem 2. SSD for Problem 1 and SSF for Problem 2 are both elaborated and include a diagram, articulation of the principles used to find intermediate variables, and clear justification for the final result. However, both the elaborated solutions SSD and SSF have two canceling mistakes that lead to the correct final answer. SSE for Problem 1 and SSG for Problem 2 are brief with no explication of reasoning and do not give away any evidence for mistaken ideas. However, the three lines of work in the brief solution SSE are also present in the elaborated SSD (similarly, the three lines of work in the brief solution SSG are also present in the elaborated solution SSF).

The researchers decided to have the TAs grade student solutions to two isomorphic problems because we wanted to observe whether the TAs used a rubric consistently when grading two similar problems. We provided TAs multiple opportunities to grade various student solutions in order to verify that TAs' grading approaches are not dependent on the context of the problem.

You are whirling a stone tied to the end of a string around in a vertical circle having a radius of 0.65 m. You wish to whirl the stone fast enough so that when it is released at the point where the stone is moving directly upward it will rise to a maximum height of 23 meters above the lowest point in the circle. In order to do this, what force will you have to exert on the string when the stone passes through its lowest point one-quarter turn before release? Assume that by the time you have gotten the stone going and it makes its final turn around the circle, you are holding the end of the string at a fixed position. Assume also that air resistance can be neglected. The stone weighs 18 N.

**FIGURE 1**. Problem 1. The correct answer is 1292 N.

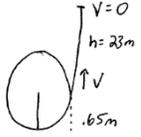

**FIGURE 2.** Student Solution D (SSD) and Student Solution E (SSE).

A friend told a girl that he had heard that if you sit on a scale while riding a roller coaster, the dial on the scale changes all the time. The girl decides to check the story and takes a bathroom scale to the amusement park. There she receives an illustration (see below), depicting the riding track of a roller coaster car along with information on the track (the illustration scale is not accurate). The operator of the ride informs her that the rail track is smooth, the mass of the car is 120 kg, and that the car sets in motion from a rest position at the height of 15 m. He adds that point B is at 5 m height and that close to point B the track is part of a circle with a radius of 30 m. Before leaving the house, the girl stepped on the scale which indicated 55 kg. In the roller coaster car the girl sits on the scale. Do you think that the story she had heard about the reading of the scale changing on the roller coaster is true? According to your calculation, what will the scale show at point B?

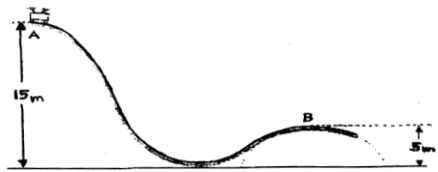

**FIGURE 3**. Problem 2. The correct answer is 180 N.

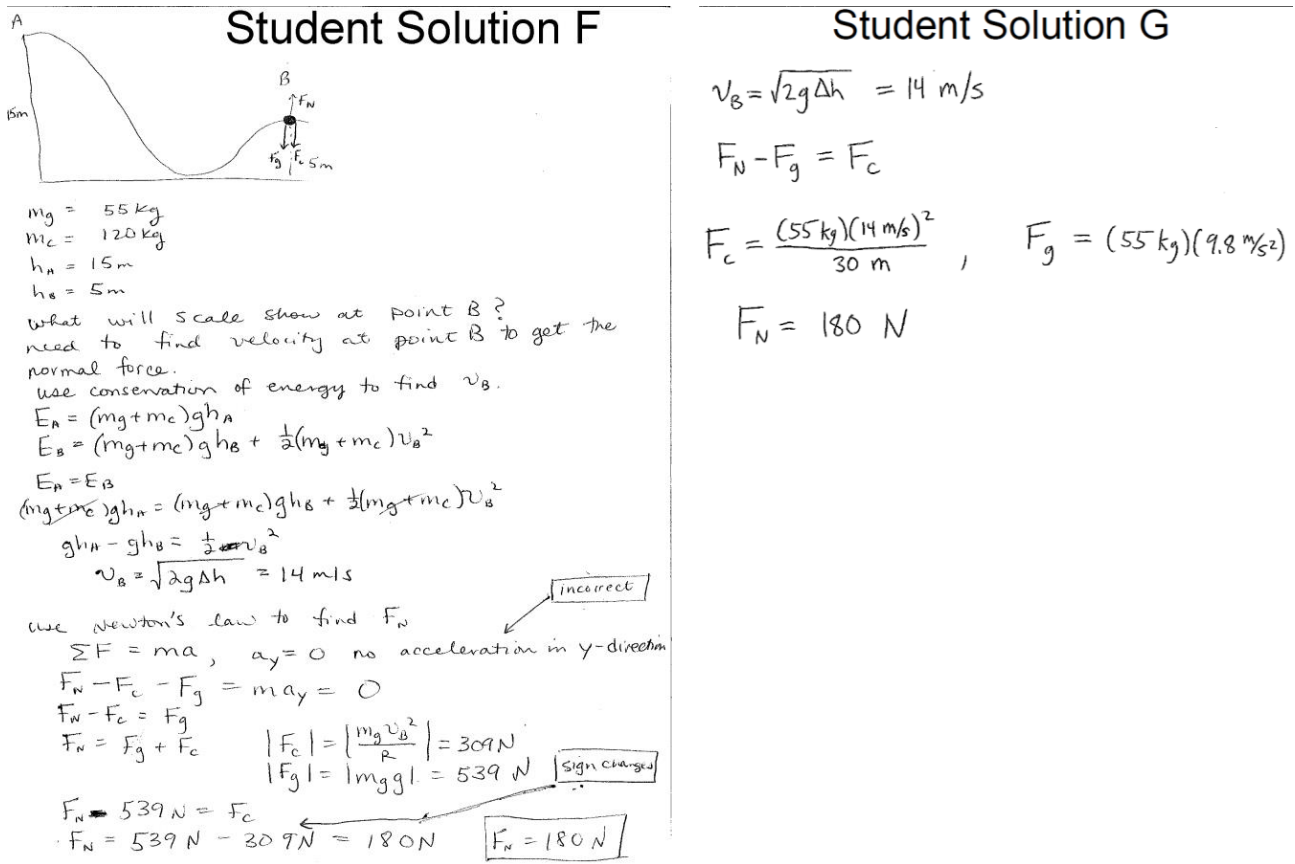

**FIGURE 4**. Student Solution F (SSF) and Student Solution G (SSG).

A standard grading rubric was developed collaboratively by four physics education researchers and iterated many times before it was implemented in this study (see Table I). The rubric emphasizes critical aspects of problem solving (e.g., invoking and justifying physics principles, evaluating of final solution, etc.) that have been found in the literature to develop problem solving skills and improve physics content knowledge [2-11]. In addition, it was designed to be general enough that it could be applied to a variety of physics problems. It is similar to the Docktor et al. rubric [39] in that it divides the grading into five separate categories: our category of problem description is similar to the Docktor et al. rubric category of "useful description," explication and justification are similar to "specific application of physics," conceptual understanding is similar to "physics approach," mathematical procedures is a rubric category in both our rubric and the Docktor et al. rubric, and problem evaluation is similar to "logical progression." The rubric in this study was designed to be more concrete in its application by dividing some of the categories into subcategories and by providing more specification of the categories. Table I also includes how an "expert" grader (e.g., an instructor who is aware that effective grading practices can help foster and support the development of problem solving skills and physics learning and has experience in grading using rubrics that weight the process of solving problem) would apply the rubric to grade the four student solutions (SSD, SSE, SSF, SSG). The "expert" scores for the four student solutions rubric scores were determined by four authors grading the four student solutions using the rubric. The initial variation in the "experts'" total scores using the rubric was within 1 point for each solution. The experts compared their grading until agreement was reached on the final "expert" scores.

### C. Sequence of grading activities within the TA professional development course

The sequence of grading activities in the TA professional development course is shown in Table II, which corresponds to the GAIQ activities focusing on grading. See Table II for a description of the homework assigned in Week 1 of the grading activity. The Week 1 homework activity elicited TAs' initial ideas about the purposes of grading and their actual grading approaches for both homework and quiz contexts. These two contexts were chosen as they were expected to trigger different grading considerations for TAs. The TAs were not given a rubric at this stage, and they were told to assume that 1) they were the instructors of the class and can structure their grading approaches to improve learning, 2) they had the authority to make grading decisions, and 3) they had told their students how they would be graded. An example response (transcribed) is shown in Fig. 5.

**TABLE I.** Rubric used to grade SSD and SSE (for Problem 1) and SSF and SSG (for Problem 2), including scores assigned to the student solutions by expert raters.

| Sample grading rubric | | | % (points) | Solution | | | |
|---|---|---|---|---|---|---|---|
| | | | | D | E | F | G |
| **Problem Description:** Evidence that the students tried to translate the problem statement into terms related to appropriate principles (2 points) (20%) | Diagram clarifying parts of the problem (1 point) | Diagram is comprehensive | +10% (+1 point) | | | 1 | |
| | | Diagram is partial | +5% (+0.5 point) | 0.5 | | | |
| | | Diagram is not present | +0% (+0 points) | | 0 | | 0 |
| | Knowns and unknowns are listed, providing evidence of an attempt to plan their problem-solving approach (1 point) | List is comprehensive | +10% (+1 point) | | | 1 | |
| | | List is partial | +5% (+0.5 point) | 0.5 | | | |
| | | List is not present | +0% (+0 points) | | 0 | | 0 |
| **Explication and justification** of the principles and concepts that are relevant to the analysis of the problem (2.5 points) (25%) | Invoking principle(s) (1.5 point) | Principles that are useful to solve the problem are invoked* (e.g., if two principles are involved, then split scoring for each) | +15% (+1.5 points) | 1.5 | | 1.5 | |
| | | Principles that are NOT useful to solve the problem are invoked* | 0% (+0 points) | | | | |
| | Justifying principle(s) (1 point) | Principles that are useful to solve the problem are justified with respect to the problem scenario* | 10% (+1 point) | 1 | | 1 | |
| | | Principles that are NOT useful to solve the problem are invoked, however they are justified with respect to the problem scenario* | 5% (+0.5 points) | | | | |
| **Conceptual understanding** (3 points) (30%) | Applying principle(s), which provide evidence that the student has an adequate understanding of the relevant principles and concepts (3 points) | Principles applied adequately* (e.g., if two principles are involved, then 15% (+1.5 point) each) | +30% (+3 points) | | | 1.5 | |
| | | Principles applied are partially correct* (w/ sign errors, missing terms, etc.) | 15% (+1.5 point) | 1.5 | 1.5 | 0.75 | 1.5 |
| **Mathematical procedures** (1 point) (10%) | Executing the solution by selecting appropriate mathematical procedures and following mathematical rules (1 point) | Algebraic procedures applied adequately | +10% (+1 point) | | 1 | | 1 |
| **Problem Evaluation:** Evidence of reflection on the problem-solving process (1.5 point) (15%) | Reasonability check of intermediate target variables and answer, e.g., checking consistency of units, limiting cases, realistic numbers, etc. (1.5 point) | Intermediate target variables and answer are *reasonable* and there is evidence of an *attempt to check the reasonability of the solution* | 20% (+2 point, extra credit for checking) | | | | |
| | | Intermediate target variables and answer are *reasonable* and there is *no evidence* of an attempt to *check the reasonability of the solution* | 15% (+1.5 point) | 1.5 | 1.5 | 1.5 | 1.5 |
| | | Intermediate target variables and answer are *unreasonable* and there *is evidence* of an attempt to *check the reasonability of the solution and/or student acknowledges that the answer is unreasonable* | 20% (+2 point, extra credit for checking) | | | | |
| | | Intermediate target variables and answer are *unreasonable* and *no acknowledgement has been made by the student* | +0% (+0 points) | | | | |
| | | Total Possible | 100% (10 points) | 6.5 | 4.0 | 8.25 | 4.0 |

*If the problem involves only one physics principle, 15% can be given for invoking the appropriate principle correctly, 10% can be given for correct justification of the principle, and 30% can be given for applying the principle adequately. If the problem involves multiple physics principles, the total percentage possible can be divided among the principles. For example, if two principles are involved then the student could get 15% for each one he or she applied adequately for a total of 30%.

See Table II for a description of the Week 2 grading activities. During the Week 2 in-class activity, two of the researchers were present in the class. One of the researchers coordinated the class work and facilitated the discussions. The group and whole-class discussions in the Week 2 – in class activity were observed and documented. As mentioned earlier, prior investigations showed that TAs' generally have productive goals for grading [21,22], i.e., that grading can serve as a learning opportunity for the students and give feedback to the instructor about students' difficulties. We aimed to build on these productive beliefs and connect TAs' initial ideas about grading with the grading rubric categories during the Week 2 in-class grading activities. In small groups, TAs discussed their grading approaches (without using a rubric) and tried to come to a consensus when grading the solution. Then, each group shared with the entire class how they graded the student solutions and any disagreements they had when grading. During the entire class discussion at the end of the class, the instructor highlighted grading approaches that promote effective problem solving and noted the disadvantages of grading which focused exclusively on correctness. The discussion also included listing the grading criteria the TAs used to grade the student solutions and then deciding as a class whether they agreed or disagreed on the appropriateness of these criteria. These criteria included listing initial information, drawing a diagram, proof of understanding, errors in physics reasoning, intermediate steps, correct units, admitting mistakes, etc. TAs were then given a rubric and there was a discussion about how the categories of the rubric aligned with their stated purposes for grading (i.e., to help students learn physics and give specific feedback to the instructor about students' difficulties). The TAs and the instructor also discussed how the rubric incorporated many of the grading criteria mentioned in the class discussion (e.g., "list" and "diagrams" as initial problem description, "proof of understanding" as explication and justification of physics principles, etc.). Each category of the rubric was explained so that TAs would understand how to apply it appropriately. A homework was assigned in Week 2 (see Table II) in which TAs graded SSD, SSE, SSF, and SSG using a rubric. The TAs were told to assume that they had distributed the rubric to their students and told them that they would be graded using the rubric.

**Table II.** Sequence of TA grading activities in the TA professional development course.

| Time | Activity |
|---|---|
| Week 1 – homework | Individually, TAs wrote an essay about the purpose of grading, graded student solutions SSD and SSE in homework and quiz contexts, and gave reasons for their final score. |
| Week 2 – in class | In small groups, TAs graded the student solutions SSD and SSE and participated in a whole class discussion about their grading approaches. The instructor and TAs discussed how many of the grading approaches mentioned by the TAs were beneficial for helping students learn effective problem-solving approaches. The instructor introduced the rubric and had a discussion with the TAs about how the rubric incorporated many of grading criteria mentioned by the TAs. |
| Week 2 – homework | Individually, TAs were asked to grade student solutions SSD, SSE, SSF, and SSG using the rubric, list pros/cons of the rubric, and describe how the rubric could be improved. |
| Week 3 – in class | In small groups, TAs discussed the scores they gave using the rubric. Then, they used the rubric again to come to a consensus on a final score in small groups. The TAs also participated in a whole-class discussion about the challenges they faced when trying to reach a consensus in the final score and gave recommendations for improvements to the rubric in order to allow different graders to be consistent in their grading. During the whole-class discussion, the instructor highlighted effective applications of the grading rubric. |
| End of semester – homework | Individually, the TAs wrote an essay about the purpose of grading, graded student solutions SSD and SSE in homework and quiz contexts, and gave reasons for their final score. |
| End of semester – in class | TAs were given copies of their Week 1 homework and asked to make comparisons between their responses from the beginning of the semester to the end of semester. The TAs were not provided a rubric at this stage. |

| Features: Solution E | Score | | Reasons: explain your reasoning for weighing the different features to result with the score you arrived at. |
|---|---|---|---|
| | Homework | Quiz | |
| The answer is correct, the approach is correct. The steps for getting $v^2=2gh$ are not written | 8 | 10 | I gave this student a lower grade on HW because I think that students have enough time to write down all steps, and they should. This answer looks like it has been written just to get a grade, not that the student was learning something while doing the HW. I think that since the approach and the answer are right, this answer gets a full grade on a quiz. |

**FIGURE 5**. One component of a sample TA's worksheet (transcribed) related to SSE which was part of the pre-class grading activity.

See Table II for a description of the Week 3 in-class grading activity. The Week 3 grading activities attempted to help TAs internalize the rubric and reflect on the pros and cons of using a rubric to grade. The small group discussions focused on TAs' agreements and disagreements about the application of the rubric and ways to improve the rubric. After these discussions, there was an entire class discussion during which each group shared how they applied the rubric and whether they were able to come to a consensus when grading using the rubric. The instructor gave positive feedback if a group of TAs noted that they used the rubric productively (i.e., used it as an expert would). The small group and whole-class discussions about the rubric aimed to help the TAs think deeply about the rubric and adapt the rubric to use in their own courses.

After the initial grading activities in weeks 1 – 3, the TAs participated in other activities in the TA training course. These activities involved, e.g., discussions of how different problem types (i.e., multiple choice problems, context rich problems, and problems that are broken into sub-problems) and different example solutions to problems can help students learn physics. TAs also were given a physics problem and asked to present the solution to the TA training class as they would in their recitations. While the TAs individually worked through the physics problems at the board in the TA training class, they were video-recorded. After they worked through the physics problem, they were asked to reflect on their teaching and also received feedback from other TAs and the instructor on their teaching approach.

See Table II for a description of the end of semester activities. These activities examined the effect of the instructional activities involving the rubric in Weeks 1-3 as well as the entire professional development course and TAs' own teaching experiences. The TAs were not given a rubric at this stage because we wanted to observe whether TAs generated a rubric on their own and the types of solution features TAs took into account when grading. The purpose of these activities were meant to help TAs reflect on how their grading did or did not change over the course of the professional development program and one semester of teaching experience.

### D. Post course interviews

After an initial analysis of the collected data, in the following semester, seven of the TAs that had been enrolled in the TA professional development course volunteered to be interviewed to provide further clarification of their stated grading beliefs (which sometimes appeared to contradict their actual grading practices). The interviews allowed the researchers to investigate whether the grading activities carried out in the TA professional development course impacted the TAs' beliefs about their grading in some manner not captured in their written responses, how they graded in actual courses for which they were TAs, and what they thought were the pros and cons of using a grading rubric. The interviewer had some pre-determined questions to ask the TAs (e.g., What in your view are the pros and cons of grading on a rubric? Have your beliefs about grading changed due to the interventions in the TA professional development course? What caused the change in beliefs?). However, the interviewer also asked additional follow-up questions on-the-spot to examine TAs' reasoning and also to give them an opportunity to clarify their written responses on the GAIQ worksheets if there were any ambiguities in their responses.

# V. RESULTS

### A. TAs' initial grading practices

We analyzed TAs' responses to the Week 1- homework activity (in which they graded SSD and SSE individually without using a rubric). Here, we discuss TAs' scores to SSD and SSE in the quiz context. The average score for the elaborated solution SSD was 7.9 and the average score for the brief solution SSE was 7.1. Seven out of the 15 TAs scored the elaborated solution SSD higher than the brief solution SSE, 6 TAs scored the brief solution SSE higher than the elaborated solution SSD, and 2 TAs gave SSD and SSE the same score. The TAs' scores in the homework context is in the Appendix in Tables I and II. Histograms of TAs' grades given to SSE and SSD are shown in Figure 1 in the appendix. We also found that the majority of the TAs graded on correct application of physics concepts and correct final answer and did not grade on problem description, evaluation, or explication of the problem solving approach. This result is similar to our previous finding [21,22].

### B. TAs' grading practices using a rubric

Here, we discuss how the TAs applied the rubric in the Week 2 - homework activity. In this activity, TAs were asked to individually grade the student solutions SSD, SSE, SSF, and SSG using the rubric, list pros/cons of the rubric, and describe how the rubric could be improved.

#### a. TAs' use of the rubric when grading elaborated solutions SSD and SSF

To investigate research questions 1, 2, and 3, we analyzed how the TAs applied the rubric when grading the student solutions. Fig. 6 shows the percentage of TAs who selected each category in the rubric when grading the elaborated solution SSD. Bars that are the same color represent categories that are usually graded in an either/or case for one of the categories. There was consensus among the TAs that principles that were useful to solve the problem were invoked (blue bar) and were justified (brown bar), and that the algebraic procedures were applied adequately (yellow bar). We note that for the yellow bar, there was no category in the rubric for "math procedures not applied adequately" so that category does not appear in Figs. 6-9. There was less consensus among TAs about whether the diagram in SSD was comprehensive or whether it was partial (red bars) and about whether the list of knowns/unknowns was comprehensive, partial, or missing (orange bars). There was also some disagreement about whether the principles used in SSD were applied adequately or whether the application was only partially correct (green bars) and whether there was evidence of a reasonability check (purple bars). Compared to an "expert" grader, the majority of TAs agreed with the "experts" in the selection of most categories for SSD. Figure 2 in the appendix shows the distribution of the scores TAs gave to the elaborated solution SSD when using the rubric to grade.

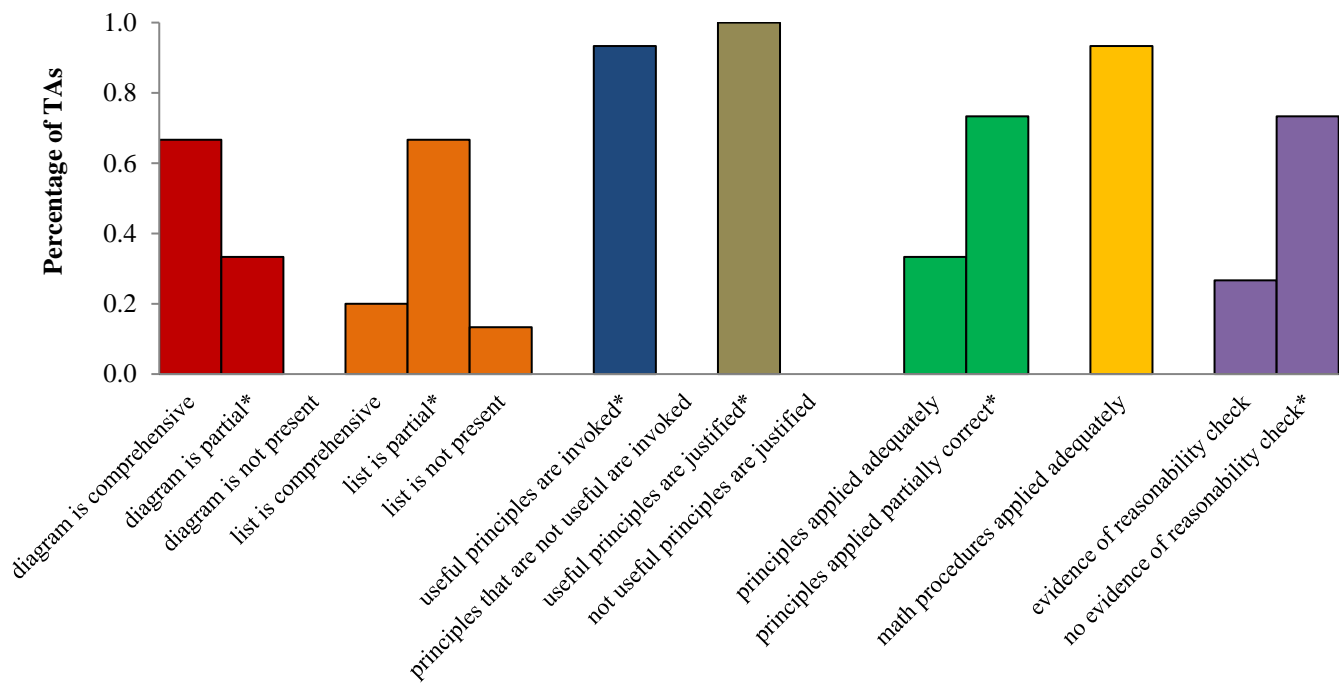

**FIGURE 6**. Percentage of TAs (N=15) who selected each rubric category when grading SSD. The categories marked with an asterisk represent the category an "expert" grader would choose when grading using the rubric. For the yellow bar, there was no category in the rubric for "math procedures not applied adequately" so that category does not appear in the figure. "Experts" did not assign points to SSD in the "math procedures applied adequately" (since there were mathematical mistakes in SSD) so there is no asterisk for the yellow bar that represents "math procedures applied adequately."

Fig. 7 shows the percentage of TAs who assigned each category in the rubric when grading the elaborated solution SSF (which was analogous to SSD). There was again consensus among the TAs that principles that were useful to solve the problem were invoked and were justified (brown bar), and that the algebraic procedures were applied adequately (yellow bar). There was less consensus among TAs about whether the principles used in SSF were applied adequately or whether the application was only partially correct (green bars), and whether or not there was evidence of a reasonability check (purple bars). Compared with an "expert" grader, the TAs mostly graded SSF as an expert would, though some TAs indicated that there was evidence of a reasonability check in the solution when that was not the case. TAs were generally consistent in applying the rubric for the analogous elaborated student solutions SSD and SSF (i.e., the percentage of TAs grading on components of the rubric for SSD and SSF are similar, as shown by comparing Fig. 6 and Fig. 7).

### b. TAs' use of the rubric when grading the brief solutions SSE and SSG

Figure 8 shows the percentages of TAs who assigned each category in the rubric when grading the brief solution SSE. There was consensus among the TAs that the diagram in SSE was not present and that the list of knowns/unknowns was also not present. TAs were also in agreement that the algebraic procedures were applied adequately, and that there was no evidence of a reasonability check. There was less consensus among TAs about whether principles that were useful to solve the problem were invoked and whether the use of those principles was justified. As with SSD and SSF, there was also disagreement about whether the principles used in SSE were applied adequately or whether the application was only partially correct. Compared to an "expert's" use of the rubric, TAs were not in agreement with an "expert" grader when selecting that "useful principles are invoked" and "useful principles are justified." There was no evidence of explicit invoking or justifying of physics in SSE, though the majority of TAs gave this solution credit for those two criteria. Figure 2 in the appendix shows the distribution of the scores TAs gave to the brief solution SSE when using the rubric to grade.

Figure 9 shows the percentages of TAs who assigned each category in the rubric when grading the brief solution SSG (which was analogous to SSE). There was again consensus among the TAs that the diagram in SSG was not present, that the algebraic procedures were applied adequately, and that there was no evidence of a reasonability check. There was less consensus among TAs about whether principles that were useful to solve the problem were invoked and whether the use of those principles was justified. As with all other student solutions, there was disagreement about whether the principles used in SSG were applied adequately or whether the application was only partially correct. As with the brief solution SSE, TAs were represent the category an "expert" grader would choose when grading using the rubric. For the yellow bar, there was no category in the rubric for "math procedures not applied adequately" so that category does not appear in the figure.

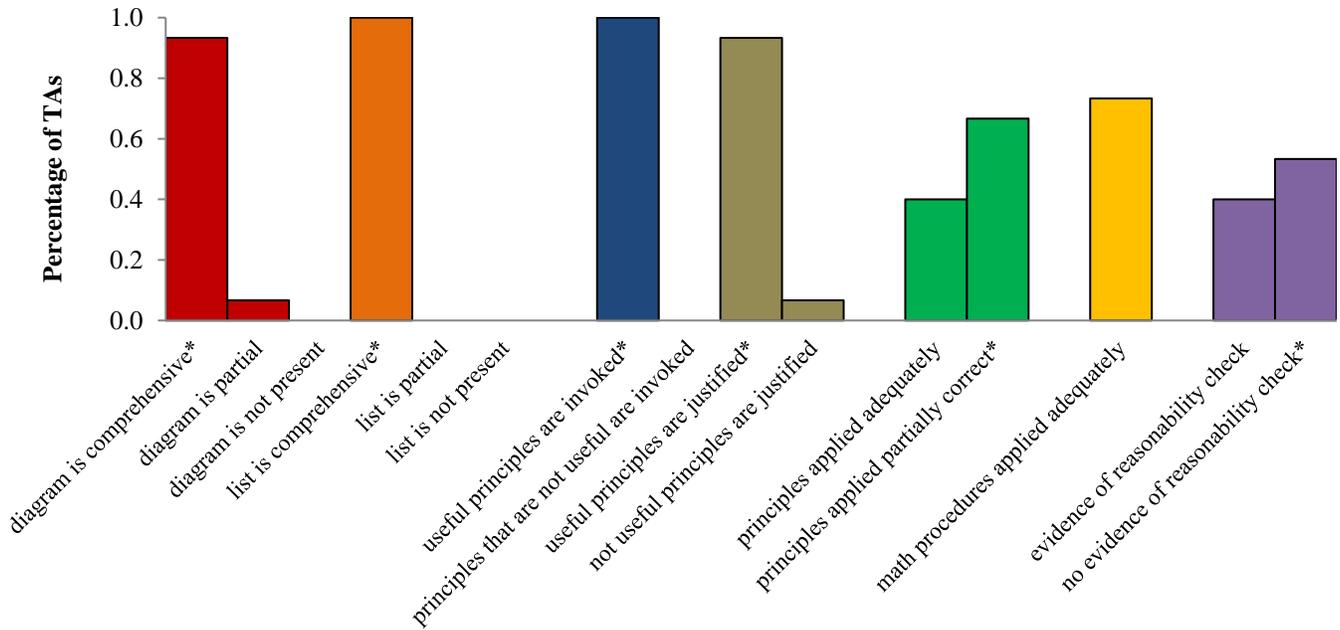

**FIGURE 7**. Percentage of TAs (N=15) who selected each rubric category when grading SSF. The categories marked with an asterisk represent the category an "expert" grader would choose when grading using the rubric. For the yellow bar, there was no category in the rubric for "math procedures not applied adequately" so that category does not appear in the figure. "Experts" did not assign points to SSF in the "math procedures applied adequately" (since there were mathematical mistakes in SSF) so there is no asterisk for the yellow bar that represents "math procedures applied adequately."

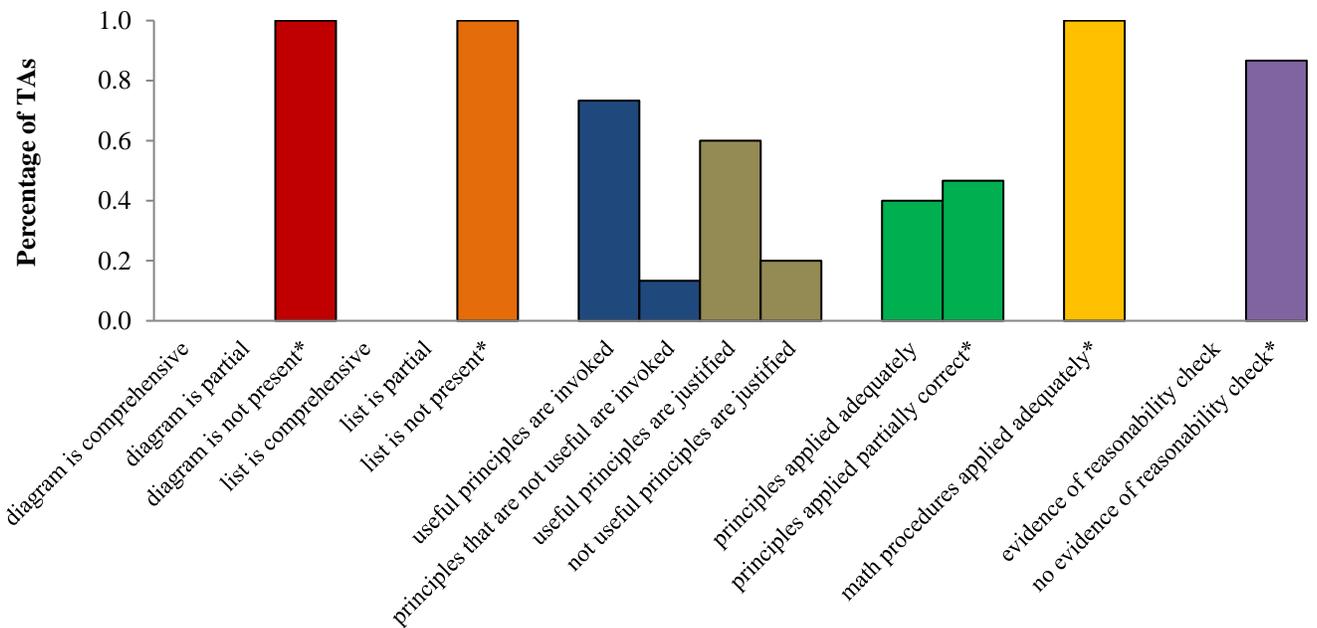

**FIGURE 8**. Percentage of TAs (N=15) who selected each rubric category when grading SSE. The categories marked with an asterisk

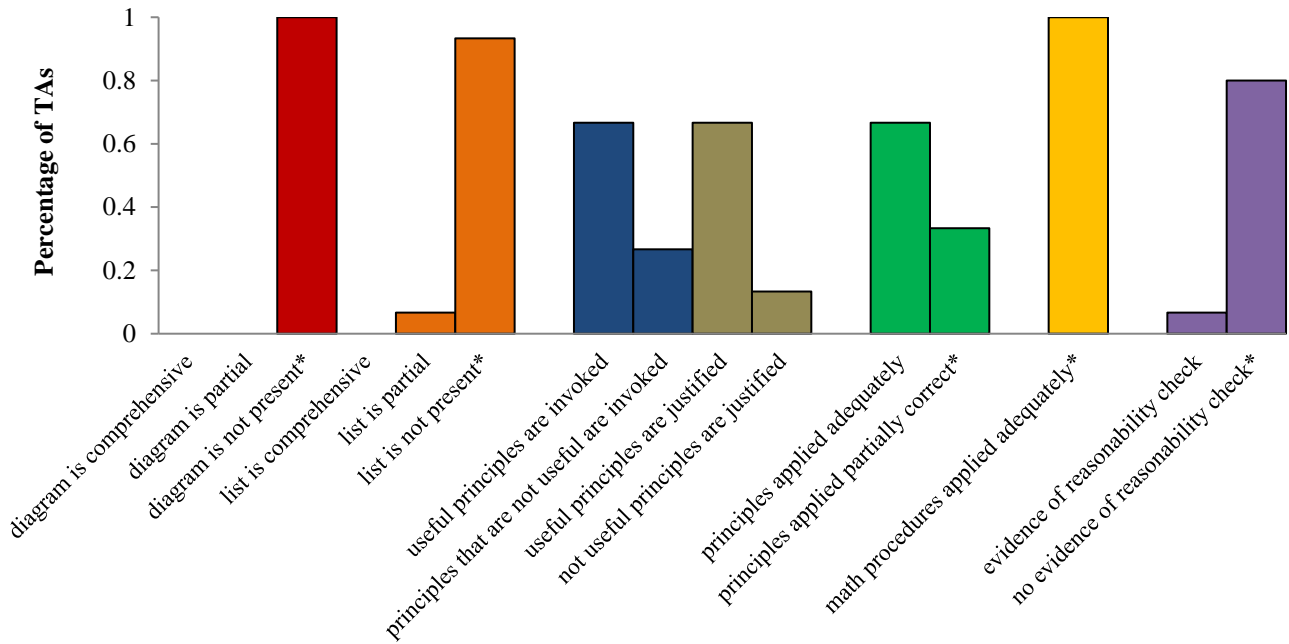

**FIGURE 9.** Percentage of TAs (N=15) who selected each rubric category when grading SSG. The categories marked with an asterisk represent the category an "expert" grader would choose when grading using the rubric. For the yellow bar, there was no category in the rubric for "math procedures not applied adequately" so that category does not appear in the figure.

again in disagreement with an "expert" grader when selecting that "useful principles are invoked" and "useful principles are justified" in the brief solution SSG, even though there was no explicit evidence of invoking or justifying in this solution. TAs were generally consistent in applying the rubric for the analogous brief student solutions SSE and SSG (i.e., the percentage of TAs grading on components of the rubric for SSE and SSG are similar, as shown by comparing Fig. 8 and Fig. 9.

## C. TAs' grading practices do not change significantly after the TA professional development course

See Table II for a description of the end of the semester grading activities. TAs' responses to the end of semester grading activities were used to investigate research question 4 (*Do TAs' grading practices change after a 15-week professional development course that included activities and extended discussions involving the benefits of using a rubric and carrying out their assigned teaching responsibilities simultaneously?*).

Table III shows the average score assigned for the elaborated solution SSD in the quiz context before the rubric was introduced in the Week 1 - homework activity, the average score assigned by TAs to SSD using the rubric individually in the Week 2 – homework activity, the score assigned by the authors to SSD using the rubric ("experts"), and the average score at the end of the semester, with standard deviations for each average score. Table IV shows the average score assigned to the brief solution SSE in the quiz context before the rubric was introduced in the Week 1 – homework activity, the average score assigned to SSE using the rubric in the Week 2 – homework activity, the score assigned by the authors to SSE using the rubric ("experts"), and the average score at the end of the semester, with standard deviations for each score. The average score assigned by TAs on the elaborated solution SSD and the brief solution SSE stayed approximately the same after one semester of a professional development course that included rubric grading activities and one semester of teaching experience. Initially in the Week 1 – homework activity, seven out of the 15 TAs scored the elaborated solution SSD higher than the brief solution SSE, 6 TAs scored the elaborated solution SSD lower than the brief solution SSE, and 2 TAs gave SSD and SSE the same score. At the end of the semester, 6 out of the 15 TAs scored the elaborated solution higher than the detailed solution, 5 TAs scored the elaborated solution lower than the brief solution, and 4 TAs gave SSD and SSE the same score. The majority of the TAs who had initially scored the elaborated solution SSD higher than the brief solution also scored SSD higher than SSE at the end of the semester. Similarly, the majority of the TAs who initially scored the brief solution higher than the elaborated solution did not change their grading practices at the end of the semester. Only one TA switched his grading practice—initially, this TA scored the brief solution SSE higher than the elaborated solution SSD and at the end of the semester the same TA scored the elaborated solution SSD higher than the brief solution. Figures 3 (a) and 3 (b) in the appendix show the distribution of the grades TAs gave to the solutions SSD and SSE at the end of the semester. We also found that, at the end of the semester, less than 40% of the TAs graded on problem description, evaluation, and explication of the problem-solving approach when grading SSD and SSE, which is aligned with our previous findings in earlier TA training courses [21,22]. If the rubric intervention had been "effective," the TAs' scores at the end of the semester should have been more aligned with the "expert" grader score and TAs would have taken into account problem description, evaluation, and explication of the problem-solving approach when grading. However, we see that TAs' scores did not change significantly from the Week-1 grading activity to the end of semester grading activity and many TAs still did not take into account these features when grading.

**TABLE III**. Average scores and standard deviations (St. Dev.) for SSD for the quiz context before using the rubric, when using the rubric to grade (score assigned by experts using the rubric is also shown), and at the end of the semester.

| SSD | Week 1 – homework activity (No rubric) | Week 2 – homework activity (Rubric) | Experts (Rubric) | End of semester – homework activity (No rubric) |
|---|---|---|---|---|
| Average | 7.93 | 7.98 | 6.50 | 8.16 |
| St. Dev. | 1.24 | 0.70 | | 1.60 |

**TABLE IV**. Average scores and standard deviations (St. Dev.) for SSE for the quiz context before using the rubric, when using the rubric to grade (score assigned by experts using the rubric is also shown), and at the end of the semester.

| SSE | Week 1 – homework activity (No rubric) | Week 2 – homework activity (Rubric) | Experts (Rubric) | End of semester – homework activity (No rubric) |
|---|---|---|---|---|
| Average | 7.07 | 6.07 | 4.00 | 7.65 |
| St. Dev. | 2.71 | 1.68 | | 3.10 |

### D. TAs' Feedback about the rubric activity via written responses, class discussions, and interviews

Since the TAs' grading practices did not change significantly after one semester of the professional development course in this study as well as one semester of teaching experience, we examined TAs' stated pros/cons of using the rubric to grade to determine if their pros/cons may have impacted whether they accepted and internalized them. Part of the assignment to use the rubric to grade SSD and SSE in the Week 2 – homework activity asked TAs to write a short essay in which they listed what they believed to be the pros and cons of using a rubric and identified changes they would make to improve the rubric. The TAs' stated pros and cons for using a rubric were coded to determine if the responses followed any trends. Based upon these trends, categories were created to describe the most common types of responses, as shown in Table V. Two researchers separately coded the responses according to the chosen categories. After coding the TAs' responses separately, the researchers compared their individual coding. The researchers initially had 85% agreement, i.e., out of all the pros and cons listed by the TAs, the researchers agreed on the coding of each pro and con 85% of the time. After discussing disagreements, the researchers reached 90% agreement. The average Cohen's kappa [58] for inter-rater reliability for all of the TAs' stated pros and cons of the rubric was 0.84. In addition, TAs also gave feedback about the pros and cons of using a rubric to grade in class discussions in the professional development course and individual interviews.

Table V shows the percentage of TAs that mention each category of pros and cons of using a rubric in their written responses (although interviews provided an opportunity for clarification in some cases). The most commonly stated drawback of using a rubric, in TAs' opinions, was that a rubric did not allow for enough flexibility when assigning scores (e.g., to give partial credit in certain cases) or they were uncomfortable in taking off points if the final answer was correct, with 53% of TAs mentioning this con of using a rubric. In particular, the TAs often felt that they should have the freedom to grade the introductory student solution in a manner they see appropriate based upon their intuition rather than being tied by a rubric. Several TAs mentioned that a rubric is too constraining and they wanted to be able to give a high score to a student whose final answer was correct even if the student did not explicate his or her problem-solving approach. In an interview, one TA noted that student solutions are too "complicated" to be graded using a rubric. He explained it further with the following statement: "the answers are not like filling in forms. They're much more interwoven and complicated than that. You cannot really say, 'okay, here we have this, so one point to that.' That's not true in the real case. So I just read it (the rubric) and got some idea out of it, but didn't really follow every instruction." Individual interviews and class discussions in the professional development class suggest that this type of feeling was common amongst other TAs as well.

In a different interview, another TA was concerned about the fact that a rubric may restrict creativity, stating, "I think the rubrics are a little too specific, because I don't think you can categorize everything just by writing a rubric. It's hard to really balance the creativity part of students going to their correct solution. So the rubric kind of is very harsh tool to say, 'okay, these are the correct solutions, and these are not.' Which, personally, I think is against the spirit of education itself." This same TA even mentioned that rubrics "make the whole class boring, make physics boring." One TA stated that even if a student has an incorrect solution, "if any student gives interesting idea in solving problems, we should give them extra points to encourage students to think." Further discussions suggest that this TA thought that by following the criteria on a rubric, all students would be forced to solve a problem using the same approach. These types of feelings about a rubric are interesting considering the fact that the rubric the TAs were provided is not constraining in terms of how students approach a physics problem or which physics principles they use to solve the problem so long as they show the problem-solving process.

**TABLE V.** Explanation of each category used for coding TAs' stated pros/cons when grading student solutions using the rubric provided, with percentage (%) of TA responses mentioning each category in their written responses.

| Code | Definition | Examples | % of TAs |
|---|---|---|---|
| (Con) Lack of flexibility/discomfort in taking off points if final answer is correct | Using the rubric leads to less flexibility when grading. The graders have less freedom to assign points the way they would like to. | --"The rubric doesn't allow for much nuance. A solution that is really good may not exactly hit the mark on every category, but the student may have still demonstrated their understanding."<br>--"Restricted marking"<br>--"Over formatting/ kill diversities (of student responses to score points, e.g., short and long solutions could both be worthy of high points if they are both correct)."<br>--"A con is that partial credit may be harder to come by (for what I want to give them points for, e.g., more points for the correct final answer)." | 53% |
| (Con) More time-consuming | Use of the rubric would require either students or graders to spend more time on the problem. | --"Forces students to spend more time on (solving each problem)<br>--"Takes more time to evaluate." | 20% |
| (Pro) Encourage students to use good practices | The rubric encourages students to use effective problem-solving strategies and practices, such as drawing a diagram and justifying their use of physics principles. | --"Encourage students to follow a procedure for problem solving."<br>--"They learn better strategies for problem solving."<br>--"This rubric favors the solutions that show explication and justification of the principles and concepts, which will help students pay more attention to linking the specific physical scenario with the physical theories." | 73% |
| (Pro) Fairness/consistency | Using the rubric makes the grading process fairer for the students. When grading with a rubric, graders are more consistent with their scores. | --"This rubric will give a standard on how to grade, it is very useful to make a just assessment."<br>--"Evaluate the exams and homework fairly."<br>--"reduced the fluctuations of a grader." | 40% |
| (Pro) Identify specific difficulties | Grading with the rubric helps the grader/instructor to identify students' specific difficulties with the material. | --"The teacher can understand at what part of the problem most students are making a mistake and he can focus on that more."<br>--"Make it easier for student to get feedback."<br>--"I do think this would be helpful for instructors, since it would be easy for an instructor to look across the grades by rubric and see where students most often lost points." | 33% |
| (Pro) Easier to grade | The rubric makes the grading process easier for the graders. | --"The pros are that it is easier to grade."<br>--"Easier for partial marking for incomplete answers."<br>--"Easier to point out mistakes." | 13% |

Some TAs were also concerned about whether the rubric would be more time-consuming, either for the students, who would be required to include details such as diagrams and justifications for their work, or for the TAs, who would be required to evaluate additional aspects of the student solutions (mentioned by about 20% of the TAs). For example, in an interview one TA said, "I think in the real world, TAs and graders don't really have much time to look at everything students write, so I think it's important to be concise and write down all that is needed and not more." This same TA also mentioned that requiring students to spend more time on the process may be unnecessary, stating: "The process is one factor, but it's not really that important… I think in most practical cases, the correct answer should be more important then (the process)." Several TAs explicitly mentioned that they had seldom been penalized for not showing the process in their own courses and did not feel comfortable taking off points if the final answer was correct.

An issue that several TAs mentioned in interviews (but not in their written list of pros/cons) was that they may not use the rubric especially in the quiz which has time constraint if they can infer student understanding from looking at a student's solution. For example, one TA stated: "When students take a quiz, I know that he's not cheating so he knows the answer, but maybe he's stressed or trying to do it really fast, so he did part of it in his mind. I'm sure that he did the right thing for the quiz so I gave him the full grade for the quiz." Another TA mentioned that he identifies with students who write brief solutions, stating: "In my past I've usually answered questions in that form [of a brief solution like SSE], so I guess I can understand what students are trying to say when they write things like that." This TA was among those that gave the brief solutions SSE and SSG credit for justifying the use of invoked physics principles when grading with the rubric even though there was no explicit evidence of justification in those student solutions.

Some TAs also mentioned in interviews (but not in their written list of pros/cons) that in their opinion, grading should only serve a summative purpose. For example, according to one interviewed TA, "it is up to the students to get something out of their solution and student learning should not depend upon the TAs' grading practices." This TA believed that assigning of points to features such as diagrams and lists of unknown variables was merely "sugar coating" the students' scores, i.e., assigning points that inflated student scores and simply helped the students get a better grade but did not help them learn physics. This TA and several others felt that a significant amount of points should be given for the correct final answer. Some other TAs also had similar views about the "triviality" of grading students on their initial qualitative analysis of the problem such as drawing a diagram. Despite class discussions in the TA professional development course, they were not convinced that any student who drew a diagram and wrote down knowns and unknowns but obtained an incorrect answer should be given any more points than another student who skipped those qualitative analysis and planning stages of problem solving and obtained an incorrect answer.

Although the belief that a student who obtains the correct answer with minimal work should not lose points seems to be deeply ingrained, individual interviews with the TAs suggest that some TAs' beliefs about grading may have been positively impacted by the grading activities alongside their teaching responsibilities, even though this change was not generally reflected in their grading practices at the end of the semester. Some TAs stated that they initially were grading based completely on their intuition but that the rubric helped them grade more fairly. For example, in an individual interview, one TA stated, "in the start when I was asked to grade these (student solutions) it was just my subjective knowledge...but when you give me a rubric I will stick to the rubric and evaluate the performance based on that. Rubrics helped me because when you have a whole class you're doing justice to all of them." Another interviewed TA stated, "(At first) I was going by my basic intuition... this whole semester was a learning curve for me, and as I progressed I learned a lot." This TA was happy that he at least knew that he could use a rubric to grade students objectively for any problem (whether he would always use a rubric to grade students for all problems was unclear). Even though there was little change in the average score TAs gave to the solutions SSE and SSD, individual interviews and class discussions indicate that at least some TAs had started to think about the impact of grading students on their problem-solving processes. For example, one TA stated, "Before taking this course I mostly just looked at the answer and if it's right then good, if it's not okay then you don't get anything, but after the course I started to know that you need to look at the process." Another TA stated, "Before the rubric I was just paying attention to small details, but after the rubric there's lots of things I have to be careful about when grading … (for example) I wasn't giving any points for diagrams." Some of these TAs also mentioned that they were gradually realizing that a brief solution does not necessarily demonstrate that the student understands the concepts. The fact that the TA professional development class in which the TAs did the grading activities was running parallel to their actual teaching helped some of the TAs at least begin to start thinking about the importance of the problem-solving process. For example, one TA stated that the grading activities in the class helped but simultaneous experience with the students in the classes he was teaching also helped, "When I (initially) see this (short solution), I think, 'he knows what he's doing.' But when I interacted with the students, I saw that sometimes they actually write things and they have NO idea what they're doing, they just know this equation and just go through it. That interaction helped me to understand that the students might sometimes not know what they're doing." He noted that after interacting with the students in the class he began to understand why the discussions in the TA professional development course regarding grading students on the process of problem solving was important.

In addition, a few TAs stated that the rubric activity affected their grading approaches in actual classroom settings. For example, one interviewed TA noted that he understands the importance of grading for the process and stated, "if I was given the chance (in my own grading), I would prepare a rubric and I would have my solutions so for each question the scores would be much more distributed (rather than all or nothing)." Another interviewed TA stated, "overall, I like this idea of breaking down the marks with a rubric, so when I'm not provided with a rubric I will try to make a reasonable breakdown in my mind and I will try to break them according to that one, so in that sense I would say I like this (rubric)." Thus, even though the grading activities with a rubric that emphasized the process of problem solving did not necessarily show discernable changes in their grading approaches at the end of the TA professional development course, discussions with the TAs suggest that at least some of them were contemplating the benefits of grading that emphasizes the problem-solving process. The fact that at least some TAs were paying more attention to grading on the process is somewhat encouraging.

## VI. DISCUSSION AND CONCLUSIONS

In this study, the cognitive apprentice framework was used as a guide in designing the grading activities in the TA professional development course. The TAs were provided opportunities to practice using the rubric and reflect with their peers and the professional development course instructor how an effective rubric can support the goals of helping students learn physics and develop effective problem solving skills. We hypothesized that the cognitive apprenticeship-inspired grading activities would provide feedback and scaffolding support to help TAs improve their grading practices and focus more on the problem solving process (as opposed to the correctness of the final answer) and require that students show evidence of understanding via the explication of the problem-solving process.

In regards to Research Question 1 (How do TAs apply the different components of the rubric that weights the problem-solving process to grade student solutions of introductory physics problems?), we found that many TAs did not use the rubric as intended to grade solutions in which the final answer was correct but the problem-solving process was not explicated. Approximately 60% of the TAs claimed that physics principles were invoked and justified appropriately in the brief solutions (SSE and SSG), even though they were not. Research Question 2 focused on whether the TAs used the rubric consistently when grading solutions to problems involving the same physics principles but having different surface features. Our findings suggest that TAs applied the rubric consistently across analogous student solutions for isomorphic problems (i.e., analogous solutions SSD and SSF in which the problem-solving process is explicated and analogous solutions SSE and SSG, in which the problem-solving process is not explicated). This consistency in grading across analogous solutions to isomorphic problems indicates that TAs may hold some prior conceptions about grading (e.g., the belief that students who have the correct final answer should not be penalized significantly) and apply these ideas consistently across different student solutions for similar types of problems. In regards to Research Question 3 (Do TAs apply the grading rubric differently than an "expert rater", e.g., physics education researchers who study problem solving?), we found that TAs generally agreed with "expert raters" in their application of the rubric for the elaborated solutions. However, TAs' use of the rubric for the brief solutions was not in alignment with "expert raters" on the brief solutions, mainly because TAs inferred that the brief solutions invoked and justified useful physics principles (even though the brief solutions did not show evidence of understanding). Research question 4 focused on the change in TAs' grading practices after a professional development course and carrying out their assigned teaching responsibilities. We find that, in general, physics graduate TAs' grading approaches did not change significantly despite feedback and scaffolding support the TAs were provided (including class discussions and reflection about how a good rubric can promote effective problem-solving approaches and aid in learning physics). Comparing the grading of SSD and SSE at the beginning of the semester to the end of the semester, there was little change in the scores given to the solutions and approximately half of the TAs gave the brief solution SSE a score higher than or equal to the elaborated solution SSD. In other words, TAs continued to give the benefit of the doubt to the student who wrote a short solution and did not penalize the student for not articulating and justifying the principles used in the solution. In regards to Research Question 5 (According to the TAs, what are the pros and cons of using a rubric to grade student solutions in introductory physics?), the TAs noted that the pros of a grading rubric involved fairness and consistency, easy identification of common student difficulties, and easiness to grade. On the other hand, TAs' perceived cons of the rubric involved lack of flexibility/discomfort in taking off points if the final answer is correct and the time required to grade student solutions using the rubric.

In summary, we find that the cognitive apprenticeship-inspired grading activities implemented in the professional development course did not result in measurable changes in TA grading practices. On a positive note, although many TAs were not comfortable taking off points from a student solution that had the correct final answer but that did not explicate the problem solving process, individual interviews with the TAs suggest that at least a few TAs' beliefs about grading may have been somewhat positively impacted by the grading activities alongside their teaching responsibilities.

### A. Study limitations

One limitation of the study is that it is a case study with only 15 TAs at a large research university. In addition, even though the TAs were told to grade the student solutions as the instructor of the course (i.e., they were asked to assume that they are in control of the class and have told their students explicitly how they will be graded), it is possible that TAs graded the student solutions as they would normally do in their class (particularly because they were serving as a TA in another instructor's class). Furthermore, the TAs did not grade actual student solutions from their own classes so the findings of this study are contextualized in a task mimicking a "real" grading situation as closely as possible (although the grading experience in the classes they were actually teaching could be very different depending, e.g., upon the course and the instructor). We also note that due to the time constraints in the TA professional development course, we were unable to spend more than three weeks on the grading activities. It is possible that, given more time and support, e.g., via discussions with the instructor and their peers, TAs' views about grading may have improved more than what we observed in this study.

### B. Possible reasons for the lack of change in TAs' grading practices in this study

TAs' beliefs and practices about teaching and learning may be difficult to change partly because of their prior experiences as students and the departmental culture and context [17,18]. For example, prior research by Goertzen et al. suggests that getting buy-in from tutorial teaching assistants is important for reforming instruction [18]. Moreover, prior research suggests that for grading, even physics faculty members may not necessarily require their students to show evidence of understanding via explication of the problem solving process (i.e., they are often hesitant to take off points if the final answer is correct but the problem-solving process is not shown) [1,33]. Since TAs may have been graded by faculty members who have only taken points off if the final answer was incorrect, TAs themselves may not value the problem-solving process when grading students' solutions and may not require their students to explicate the problem solving process. Furthermore, if a TA's supervising faculty member does not discuss these issues with the TA and insist that the TA grades a student solution on the problem-solving process, TAs may not internalize the benefit of using a grading rubric that appropriately weights the problem-solving process. In the present study, interviews with 20 faculty members in the same physics department as the TAs indicate that most of them do not require TAs to use rubrics when grading homework or quizzes. Faculty members who required that TAs use rubrics when grading exams mentioned that rubrics for exam grading ensure fairness in grading, but not a single faculty member mentioned that a consistent use of a good rubric can help students learn physics and develop effective problem solving approaches. Furthermore, TAs' coursework responsibilities and research responsibilities may limit the time they may want invest in changing their teaching practices. These factors may influence the extent to which TAs buy-in to using a grading rubrics.

In the following section, we elaborate on some possible reasons for why significant changes in TAs' grading practices from the beginning to the end of the semester were not observed after the rubric activities focusing on the problem-solving process in the TA professional development course. We intend to test these in detail in future research.

**1. Some TAs were uncomfortable requiring students to show evidence of understanding partly because of their own past experiences as students as well as departmental contexts.** TAs' written work and class discussions suggest that even at the end of the 15-week semester, some TAs continued to infer information from introductory physics student solutions which was not explicitly stated. In fact, even when TAs were given a rubric which included criteria for invoking and justifying physics principles, a majority of TAs were willing to give the brief solutions SSE and SSG (for Core Problems 1 and 2, respectively) credit for justifying principles even though those solutions did not contain any form of explicit justification. As noted, some TAs explicitly noted that they were uncomfortable requiring students to show evidence of understanding. They explained their opinion by stating that they themselves wrote brief solutions and expected to get full scores if the final answer was correct in most of their own course work. Since most TAs may not have been penalized for not showing proof of understanding in their solutions in their courses, they may empathize with their students for using a similar approach. They may read between the lines of a student's solution and assume that they understand what their students know when their solutions do not show the problem-solving process but have the correct final answer. Individual interviews with some of the TAs confirms this hypothesis.

In addition, polling of 20 faculty members at the same university suggests that except for exams, very few physics instructors require that their TAs use rubrics to evaluate their students on a regular basis in homework and quizzes. The fact that few physics instructors require their TAs to use rubrics may also partly explain why TAs felt uncomfortable using a rubric to penalize students for not explicating the problem-solving process and demonstrating understanding. The social context of the cognitive apprenticeship model may not have been successful in changing TAs' views about grading, which appear to be highly ingrained. It appears that the social context of the physics department in which this study was conducted may not have supported the goals of the professional development course (e.g., helping TAs develop effective grading practices). It is possible that TAs did not discern any problems with their grading practices partly because their supervising instructors were themselves not aware of the benefits of grading using a rubric and did not discuss these issues with the TAs. Thus, the discussions with peers and the instructor in the TA professional development course may not have been sufficient to change TAs' grading practices.

**2. The professional development course may not have effectively taken into account TAs' own prior experiences sufficiently when the TAs were graded as students.** Some TAs noted that when they looked at their own graded solutions, they were mainly concerned about whether the final answer was correct or incorrect and would only try to figure out where they went wrong and fix their mistakes if their final answer was incorrect. They often drew parallels between what they would do when they performed poorly and what their students should do if they performed poorly. Some TAs stated that if the students did not arrive at the correct final answer, they should try to figure out where they went wrong and learn from their mistakes. Some TAs noted that if they do not get the correct answer, they generally try to learn from their mistakes. Therefore, they focused on whether their student's final answer was correct or incorrect when grading and stated that students should figure out their mistakes on their own once they realize that their answer is incorrect. These TAs did not expect detailed solutions from students that explicate the problem-solving process nor did they grade on the problem-solving process because they felt that getting an incorrect final answer should motivate students to learn how to solve the problem correctly. The TAs' belief that students should learn from their mistakes was productive. However, TAs did not realize that undergraduate students do not necessarily use their mistakes as a learning opportunity. Moreover, requiring students to show evidence of understanding and grading on the problem-solving process using a grading rubric can serve as formative assessment tool, i.e., it can provide focused feedback to the students so that they can determine their strengths and weaknesses in their problem-solving skills and

content knowledge. Future professional development courses can build on TAs' productive belief that students should learn from their mistakes.

**3.  The professional development course may not have been interactive enough.** Another possible reasons for why many TAs' grading practices did not change significantly as a result of the activities involving the grading rubric (emphasizing the problem-solving process) was that they were not involved in the development of the rubric and may not have liked the rubric. Moreover, they were not asked to use the same rubric for the course they were teaching that semester. Due to the fact that they were not involved in the development of the rubric and were not asked to use the same rubric for the classes they were teaching, they may not have engaged with the rubric effectively or did not deeply contemplate the class discussions about why such a rubric is useful. Some TAs claimed that rubrics are too restrictive, either for the graders or for the students. In individual interviews, some TAs did not seem to acknowledge that the rubric they were given can account for many different methods for solving the problem. Some of them noted that they did not want to penalize students who had not explicated the problem-solving process but had the correct final answer so they did not like the rigidity of the rubric. In their view, those students who had the correct final answer knew how to solve the problem and should not be penalized for not showing their work. Therefore, they often largely ignored the rubric in such cases (when a short solution with the correct final answer was provided).

Other TAs noted that, in general, they did not like rubrics that weighed the problem-solving process more heavily than the final answer because they felt that such rubrics give extra points to students for things that are unimportant and do not show understanding (e.g., drawing a diagram). Some TAs were not convinced that writing such detailed solutions, which had diagrams or known and unknown variables written explicitly, can help students become better at problem solving. They felt that assigning points to features such as diagrams and lists of unknown variables was merely inflating the students' scores, i.e., assigning points that simply helped the students get a better grade but did not help them learn physics or develop good problem-solving skills. These TAs felt that significantly more points should be given for the correct final answer because the purpose of grading was to see if the students knew how to solve the problem correctly and arrive at the final answer. Other TAs stated that grading using a rubric is too time consuming for the students (because they have to spend more time writing down their process and explanations) and the graders (because they have to spend more time grading on the process and explanations as opposed to only checking that the answer is correct). Therefore, they preferred to use their intuition to grade rather than using the rubric. In order to get buy-in, the TAs who claimed that they did not like the rubric given in the professional development course for the reasons mentioned above may need more interactivity in generating ideas about how to grade students when their problem-solving process is not explicated and in creating a rubric that takes these ideas into account.

**4.  Some TAs may have remained in a state of cognitive conflict in terms of their grading practices.** It is also possible that some TAs were impacted by the grading rubric intervention but this impact was not reflected in their grading practices at the end of the semester. Some TAs may have been in a state of cognitive conflict and it was challenging for them to assimilate what they had learned in the TA professional development course with their views about grading that they had held for a long time as students. Similar findings have been reported in the context of learning rules, e.g., for balancing, in which students have difficulty taking into account the impact of both lever arm and the weights hanging from the two sides [59]. It was found that the students were in a "mixed" state even after several rounds of intervention and sustained intervention was needed to help them internalize the rules [59]. TAs, in general, seemed unfamiliar with the concept of using a rubric that focuses on process (except to give partial credit to students for fairness) and found it difficult to accept and apply what the professional development course emphasized (i.e., that solutions that did not explicate the problem-solving process should be penalized). After being exposed to the rubric and discussing the pros/cons of the rubric, TAs could either dismiss the rubric completely, accept and internalize the rubric, or remain in a "mixed" state [59]. TAs who were in a "mixed" state may recognize the formative benefits of grading using the rubric but did not necessarily resolve to use the rubric when grading (especially when explication of the problem-solving process was missing but the final answer was correct). Our written and oral data from class discussions and individual interviews suggest that while some TAs may have dismissed the rubric provided (or the idea of a rubric altogether, preferring to use their intuition alone to grade), others may have needed more time to internalize the rubric since it was an unfamiliar grading tool and penalizing students for the process when the final answer was correct was too discomforting for them.

Even though changes in TAs' grading practices were not apparent, some TAs indicated in interviews that they were still contemplating the value of grading students for the process of problem solving as a result of the rubric activity several weeks after the TA professional development course was over. These TAs may need more time and more exposure to reflect on the benefits of the rubric. It is possible that with more time and exposure to reflect on the formative benefits of grading using a rubric that explicates the problem-solving process, they would realize that rubrics that focus on the process of problem-solving can help students develop effective problem-solving approaches and learn physics.

## C. Implications

Leaders of the professional development courses for physics graduate TAs and physics education researchers can take advantage of the findings of this study. Future studies can build on this research and investigate strategies to get buy-in from the TAs so that they require students to show evidence of understanding by consistently using a rubric that appropriately weights the problem-solving process. Helping TAs grade students' solutions on the process of problem solving requires extended time, discussion, support, feedback and practice. The professional development courses can allow more support for the TAs to internalize how grading can be used for formative assessment, i.e., that grading using a well-designed rubric can support students in developing better problem solving practices and learning physics better. It may be helpful to have the TAs participate in the development of a rubric in order to help them feel more engaged in the grading activity. TAs can also be encouraged to use the rubric to grade students' solutions in the recitations that they are teaching in a particular semester and track the changes in their students' problem solving practices and learning over the semester. Over time, this practice may allow the TAs to observe how a good rubric can make grading more objective and encourage students to adopt effective problem-solving strategies. Professional development courses can also take into account TAs' productive belief that their students should learn from their mistakes and help TAs reflect on how grading using a rubric can provide students focused feedback that helps them repair their knowledge structure and develop better problem-solving skills. TAs might also benefit from being asked to think about situations in which they were given limited feedback vs. detailed feedback in their own graded solutions and reflect on how the detailed feedback helped them learn physics and develop better problem-solving skills. In addition, physics instructors who supervise graduate TAs can collaborate with their TAs in creating grading rubrics since the departmental context may at least partly account for whether TAs buy into consistently using a rubric. It is possible that the TAs will then begin to require students to show evidence of understanding via explication of the problem-solving process and view grading using a rubric as a means to support student learning.

## ACKNOWLEGEMENTS

We thank the members of the physics education research group at the University of Pittsburgh as well as the TAs involved in this study.

## APPENDIX

**TABLE I**. Average scores and standard deviations (St. Dev.) for SSD for the homework context before using the rubric, when using the rubric to grade (score assigned by experts using the rubric is also shown), and at the end of the semester.

| SSD | Week 1 – homework activity (No rubric) | Week 2 – homework activity (Rubric) | Experts (Rubric) | End of semester – homework activity (No rubric) |
|---|---|---|---|---|
| Average | 7.40 | 7.98 | 6.50 | 7.21 |
| St. Dev. | 1.30 | 0.70 | | 1.49 |

**TABLE II**. Average scores and standard deviations (St. Dev.) for SSE for the homework before using the rubric, when using the rubric to grade (score assigned by experts using the rubric is also shown), and at the end of the semester.

| SSE | Week 1 – homework activity (No rubric) | Week 2 – homework activity (Rubric) | Experts (Rubric) | End of semester – homework activity (No rubric) |
|---|---|---|---|---|
| Average | 6.00 | 6.07 | 4.00 | 6.13 |
| St. Dev. | 3.16 | 1.68 | | 2.85 |

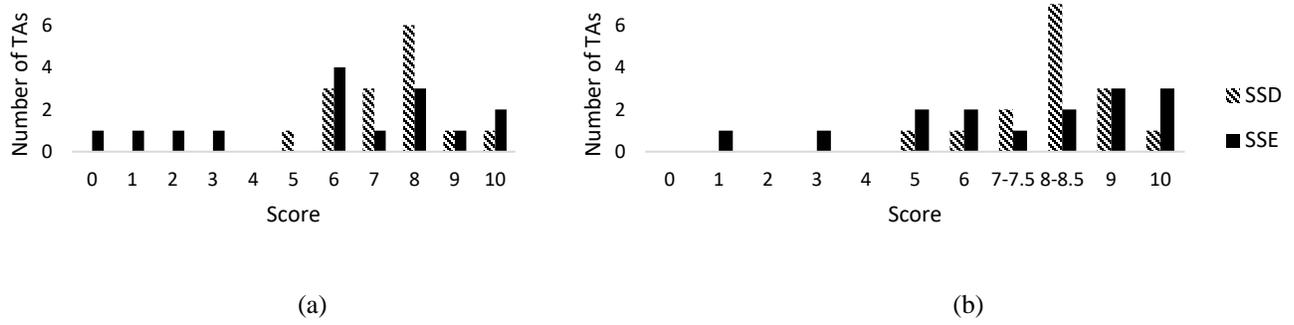

**Figure 1.** Histogram of TAs' initial scores on the elaborated solution SSD and the brief solution SSE in the homework (a) and the quiz (b) before using a grading rubric.

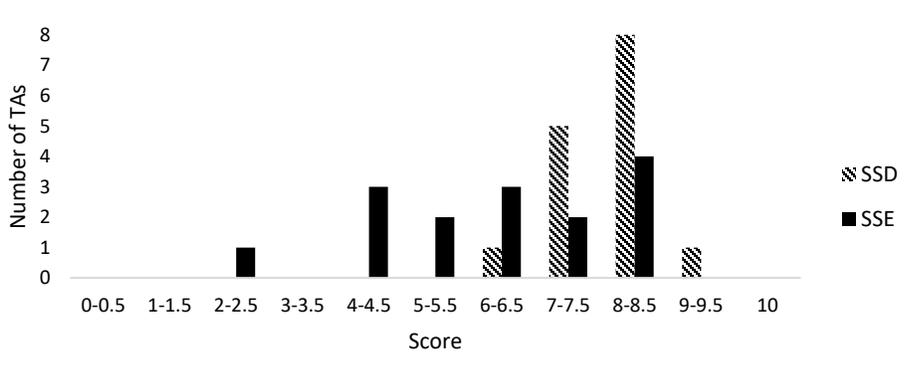

**Figure 2.** Histogram of TAs' scores on the elaborated solution SSD and the brief solution SSE when using a grading rubric.

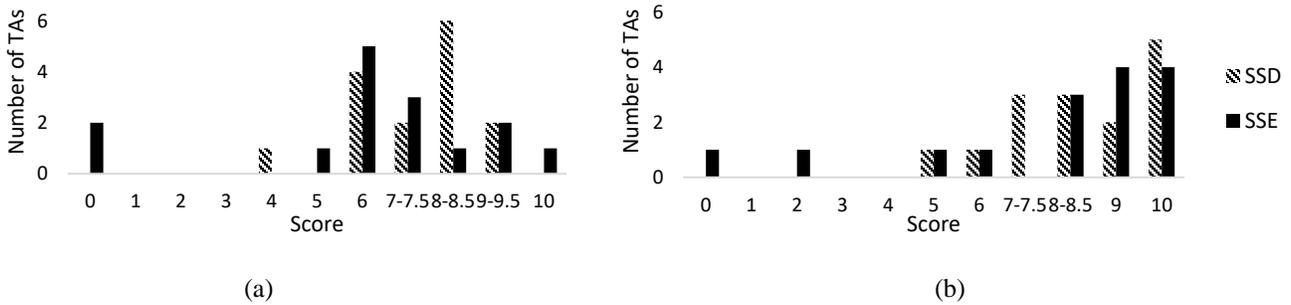

**Figure 3.** Histogram of TAs' end-of-semester scores on the elaborated solution SSD and the brief solution SSE in the homework (a) and the quiz (b).